# A smoothed particle hydrodynamics method for evaporating multiphase flows

Xiufeng Yang*, and Song-Charng Kong**

*Department of Mechanical Engineering, Iowa State University, Ames, IA 50011, USA*

* xyang@iastate.edu

** kong@iastate.edu

**Abstract**

Smoothed particle hydrodynamics (SPH) method has been increasingly used for simulating fluid flows, however its ability to simulate evaporating flow requires significant improvements. This paper proposes an SPH method for evaporating multiphase flows. The present SPH method can simulate the heat and mass transfers across the liquid-gas interfaces. The conservation equations of mass, momentum and energy were reformulated based on SPH, then were used to govern the fluid flow and heat transfer in both the liquid and gas phases. The continuity equation of the vapor species was employed to simulate the vapor mass fraction in the gas phase. The vapor mass fraction at the interface was predicted by the Clausius-Clapeyron correlation. A new evaporation rate was derived to predict the mass transfer from the liquid phase to the gas phase at the interface. Because of the mass transfer across the liquid-gas interface, the mass of an SPH particle was allowed to change. New particle splitting and merging techniques were developed to avoid large mass difference between SPH particles of the same phase. The proposed method was tested by simulating three problems, including the Stefan problem, evaporation of a static drop, and evaporation of a drop impacting on a hot surface. For the Stefan problem, the SPH results of the evaporation rate at the interface agreed well with the analytical solution. For drop evaporation, the SPH result was compared with the result predicted by a level-set method from literature. In the case of drop impact on a hot surface, the evolution of the shape of the drop, temperature, and vapor mass fraction were predicted.

flow

## I. INTRODUCTION

Because evaporation is encountered in many engineering applications, such as fuel droplets in engines, liquid sprays, and material processing [1-5], a numerical method to accurately predict liquid evaporation is of great importance. Common engineering models for predicting droplet evaporation assume that the liquid droplet is a point source with homogeneous properties [1-4]. The primary concern of these models the mass transfer rate without consideration of the gradient in the droplet or the liquid-gas interface. While such models are useful in engineering applications, advanced numerical methods are needed to reveal the details of the evaporation process.

The dynamics of evaporating flows involves phase change and energy transfer at the liquid-gas interface, diffusion of vapor species in the gas phase, and multiphase flows with sharp interfaces. Because of the complexity of the evaporation problem, it is challenging to detailed numerical simulation. The main numerical challenges in simulating evaporating flows include the treatment of phase change and the sharp discontinuity of fluid properties at the liquid-gas interface. Phase change due to evaporation causes mass transfer from one phase to another phase. The discontinuity at the liquid-gas interface, of variables such as density ratio, also leads to numerical difficulties.

Several numerical methods to address the challenges in modeling the details of evaporating flows have been developed in recent years. Tanguy et al. [6] presented a numerical method using both the level-set method and the ghost fluid method to capture the interface motion and to handle conditions at the interface. Safari et al. [7, 8] developed a lattice Boltzmann method (LBM) for simulating the phase change of multiphase flows with evaporation. Nikolopoulos et al. [9] investigated the evaporation process of n-heptane and water droplets impinging on a hot surface using the finite volume method coupled with the volume of fluid (VOF) method. Strotos et al. [10] studied the evaporation of water droplets depositing on a heated surface at low Weber numbers using VOF.

The intent of this work is to provide a numerical method, based on smoothed particle hydrodynamics (SPH), to simulate multiphase flows with evaporation. The SPH method is a Lagrangian mesh-free particle method. In SPH, a continuous fluid is discretized using SPH particles, which carry physical properties, such as mass, density, pressure, viscosity, and velocity. Since SPH is a mesh-free method, a smoothing kernel is introduced to connect the neighboring particles. The variables and their spatial derivatives are discretized in summations over particles. The SPH method was originally proposed by Lucy [11] and Gingold and Monaghan [12] for astronomy problems. Since then SPH has been applied to a wide range of problems [13-15]. In recent years, the SPH method was extended for phase change flows. By using the van der Waals (vdW) equation of state, Nugent and Posch [16] applied SPH for modeling vdW fluid drop surrounded by its vapor. Their numerical results showed that there was more vapor around the drop at higher temperature. Using SPH with vdW equation of state, Sigalotti et al. [17] simulated the rapid evaporation and explosive boiling of a vdW liquid drop. Ray et al. [18] applied the vdW-SPH method to study the liquid-vapor equilibrium of the vdW fluid. Das and Das [19] proposed a model based on SPH to describe gas-liquid phase change by introducing pseudo particles of zero mass. However, the previous phase change SPH methods consider the interaction between the liquid and its vapor, but do not consider the effect of the concentration of the vapor species in the gas phase on evaporation and the diffusion of the vapor species in the gas phase. Therefore, the ability of SPH to simulate evaporation needs further improvement.

In the classical SPH method, the mass of an SPH particle is constant, i.e., the mass of an SPH particle does not change during simulation. In the SPH method developed for this study, the SPH particles near the interface are allowed to change their mass to model the process of evaporation at the interface. The rate of mass change of SPH particles due to evaporation depends on the vapor mass fraction in the gas phase and the saturated vapor mass fraction at the interface. The saturated vapor mass fraction can be predicted by the Clausius-Clapeyron correlation. During the process of evaporation, the mass of a liquid SPH particle at the interface increases, while the mass of a gas SPH particle decreases. To constrain the mass of individual SPH particles, a particle will split into smaller particles if its mass is large enough or merge into a neighbor particle if its mass is small enough.

The rest of this paper is organized as follows. Governing equations are given in Sec. II, including the derivation of evaporation rate. Sec. III provide the numerical method, including the SPH formulations for liquid-gas interface and evaporation rate, and the particle splitting and merging technique. The numerical method is tested in Sec. IV by three different numerical examples. Then the paper ends with conclusions in Sec. V.

## II. GOVERNING EQUATIONS

The conservation equations of mass, momentum and energy are used to describe the transport of both the liquid phase and gas phase. These equations are expressed in the Lagrangian form.

$$\frac{d\rho}{dt} = -\rho \nabla \cdot \boldsymbol{u} \tag{1}$$

$$\frac{d\boldsymbol{u}}{dt} = \boldsymbol{g} - \frac{1}{\rho}\nabla p + \frac{\mu}{\rho}\nabla^2 \boldsymbol{u} \tag{2}$$

$$\frac{dT}{dt} = \frac{1}{\rho C_\mathrm{p}} \nabla \cdot (\kappa \nabla T) \tag{3}$$

Here $\rho$ is the fluid density, $\boldsymbol{u}$ is the fluid velocity, $p$ is the fluid pressure, $\mu$ is the dynamic viscosity, $T$ is the temperature, $C_\mathrm{p}$ is the specific heat at constant pressure, $\kappa$ is the thermal conductivity, and $\boldsymbol{g}$ is the gravitational acceleration. Note that in this paper the production of thermal energy by viscous dissipation is not considered in the energy equation because of its relatively small magnitude [6, 7, 20].

The following equation of state is used to calculate pressure

$$p = c^2(\rho - \rho_\mathrm{r}) + p_\mathrm{r} \tag{4}$$

where $c$ is a numerical speed of sound, $\rho_\mathrm{r}$ is a reference density and $p_\mathrm{r}$ is a reference pressure.

At the liquid-gas interface, the process of phase change due to evaporation will cause mass and energy transfer. Thus, the continuity and energy equations, Eqs. (1) and (3), at the liquid-gas interface are modified as

$$\frac{d\rho}{dt} = -\rho \nabla \cdot \boldsymbol{u} + \dot{m}''' \tag{5}$$

$$\frac{dT}{dt} = \frac{1}{\rho C_\mathrm{p}} \nabla \cdot (\kappa \nabla T) - \frac{h_\mathrm{v}}{\rho C_\mathrm{p}} \dot{m}''' \tag{6}$$

where $\dot{m}$ is the mass evaporation rate across the interface while $\dot{m}'''$ is the volumetric mass evaporation rate, and $h_\mathrm{v}$ is the latent heat of vaporization.

In order to obtain the mass fraction field of the vapor species in the gas phase, the continuity equation of the vapor species needs to be solved,

$$\frac{dY}{dt} = \frac{\nabla \cdot (\rho D \nabla Y)}{\rho} \tag{7}$$

where $Y$ is the vapor mass fraction and $D$ is the mass diffusivity of the vapor.

The governing equations listed above are not closed. An equation to calculate the mass evaporation rate is needed. A couple of equations to describe the evaporation rate have been used in the mesh-based methods, however, they cannot be directly used within the SPH method, because there are no mesh in SPH. Therefore, a new equation for evaporation rate that can be used in SPH needs to be derived.

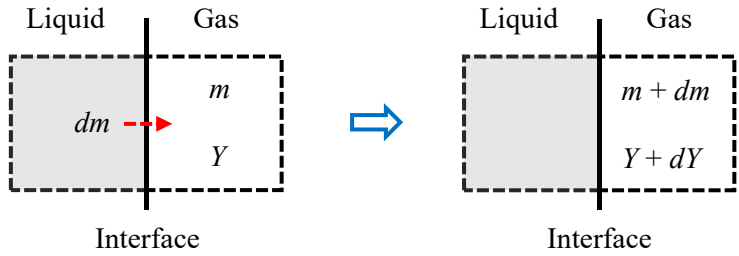

FIG. 1. Schematic of mass transfer from a liquid element to a gas element due to evaporation.

Figure 1 shows the mass transfer process at the liquid-gas interface due to evaporation. The initial total mass and the vapor mass fraction of the gas element are $m$ and $Y$, respectively. A mass $dm$ is transferred across the liquid-gas interface due to evaporation. As a result, the total mass and the vapor mass fraction of the gas element become $m+dm$ and $Y+dY$, respectively. Based on the conservation of vapor mass, we have

$$mY + dm = (m+dm)(Y+dY). \tag{8}$$

The following equation can be obtained by neglecting the second order infinitesimal term $dmdY$ in the above equation.

$$dm = \frac{m}{1-Y} dY \tag{9}$$

Then the following rate equation can be obtained.

$$\frac{dm}{dt} = \frac{m}{1-Y} \frac{dY}{dt} \tag{10}$$

Substituting Eq. (7) in the above equation yields

$$\frac{dm}{dt} = \frac{m\nabla \cdot (\rho D \nabla Y)}{\rho(1-Y)}. \tag{11}$$

Note that $\dot{m} \equiv dm/dt$ and $V = m/\rho$, Eq. (11) can be written as

$$\dot{m} = \frac{V\nabla \cdot (\rho D \nabla Y)}{1-Y}. \tag{12}$$

The volumetric mass flux can be calculated as

$$\dot{m}''' = \frac{\dot{m}}{V} = \frac{\nabla \cdot (\rho D \nabla Y)}{1-Y}. \tag{13}$$

In order to obtain the mass transfer rate across the interface, the vapor mass fraction at the interface needs to be defined. By assuming that equilibrium exists between the liquid and gas phases at the interface, the vapor mass fraction at the interface is equal to the saturated vapor mass fraction. Both can be related to the saturated vapor molar fraction as [6, 7]

$$Y_s = \frac{X_s M_v}{(1-X_s)M_g + X_s M_v} \tag{14}$$

where $Y_s$ is the saturated vapor mass fraction, $X_s$ is the saturated vapor molar fraction, $M_v$ is the molar mass of the vapor, and $M_g$ is the molar mass of the dry gas (excluding the vapor species).

The saturated vapor molar fraction, $X_s$, can be related to the saturated vapor pressure as [21]

$$X_s = \frac{p_s}{p_{ag}} \tag{15}$$

where $p_s$ is the saturated vapor pressure, $p_{ag}$ is the ambient gas pressure (including the vapor species). Then the saturated vapor molar fraction can be estimated by integrating the Clausius-Clapeyron equation [6, 21]

$$X_s = \frac{p_s}{p_{ag}} = \exp\left[-\frac{h_v M_v}{R}\left(\frac{1}{T_s} - \frac{1}{T_B}\right)\right] \tag{16}$$

where $R$ is the ideal gas constant, $T_s$ is the interface temperature, $T_B$ is the liquid boiling temperature at the ambient gas pressure condition.

## III. NUMERICAL METHODS

### A. Basic formulations of the SPH method

In SPH, the value of a function $f(r)$ at point $r_a$ can be approximated using the following integration

$$f(\boldsymbol{r}_a) \approx \int f(\boldsymbol{r}) W(\boldsymbol{r}_a - \boldsymbol{r}, h) dV \tag{17}$$

where $W$ is a kernel function and $dV$ is a differential volume element. The parameter $h$ is referred to as a smoothing length, which determines the size of the integral domain. In this paper, the following hyperbolic-shaped kernel function in two-dimensional space is used [22, 23]

$$W(s,h) = \frac{1}{3\pi h^2} \begin{cases} s^3 - 6s + 6, & 0 \leq s < 1 \\ (2-s)^3, & 1 \leq s < 2 \\ 0, & 2 \leq s \end{cases} \tag{18}$$

where $s = r/h$. This kernel function can avoid the so-called tensile instability [24] that may occur in fluid simulations using SPH method [22, 23].

In the SPH method, a continuous fluid is discretized into a set of SPH particles. These particles also have physical properties, such as mass $m$, density $\rho$, velocity $\boldsymbol{u}$, and viscosity $\mu$. Then the integration of Eq. (17) is discretized in particle summation as follows.

$$f(\boldsymbol{r}_a) \approx \sum_b f(\boldsymbol{r}_b) W(\boldsymbol{r}_a - \boldsymbol{r}_b, h) \frac{m_b}{\rho_b} \tag{19}$$

The derivatives of a function can also be discretized into particle summation. For example, the gradient of function $f$ can be obtained by differentiating the kernel in Eq. (19),

$$\nabla f_a = \sum_b \frac{m_b}{\rho_b} f_b \nabla_a W_{ab} \tag{20}$$

where $\nabla_a W_{ab}$ denotes the gradient of $W$ taken with respect to the coordinates of particle $a$. Note that in SPH, a derivative can be discretized into different summation forms [25, 26].

### B. SPH formulations for single phase fluid

By applying the particle summation, the governing equations, Eqs (1), (2), (3) and (7), can be replaced by the following SPH particle equations.

$$\frac{d\rho_a}{dt} = \sum_b m_b (\bm{u}_a - \bm{u}_b) \cdot \nabla_a W_{ab} \tag{21}$$

$$\frac{d\bm{u}_a}{dt} = \bm{g} - \sum_b m_b \left( \frac{p_a + p_b}{\rho_a \rho_b} + \Pi_{ab} \right) \nabla_a W_{ab} + \sum_b \frac{m_b (\mu_a + \mu_b)(\bm{r}_a - \bm{r}_b) \cdot \nabla_a W_{ab}}{\rho_a \rho_b (r_{ab}^2 + \eta)} (\bm{u}_a - \bm{u}_b) \tag{22}$$

$$\frac{dT_a}{dt} = \frac{1}{C_p} \sum_b \frac{m_b (\kappa_a + \kappa_b)(\bm{r}_a - \bm{r}_b) \cdot \nabla_a W_{ab}}{\rho_a \rho_b (r_{ab}^2 + \eta)} (T_a - T_b) \tag{23}$$

$$\frac{dY_a}{dt} = \sum_b \frac{m_b (\rho_a D_a + \rho_b D_b)(\bm{r}_a - \bm{r}_b) \cdot \nabla_a W_{ab}}{\rho_a \rho_b (r_{ab}^2 + \eta)} (Y_a - Y_b) \tag{24}$$

Here the term $\eta = 0.01h^2$ is added to prevent the singularity when two particles are too close to each other [25]. Note that Eq. (24) is valid for the gas phase SPH particles. A gas SPH particle has a property of vapor mass fraction $Y$, thus there are no particles that only represent vapor species.

In Eq. (22), $\Pi_{ab}$ is the artificial viscosity proposed by Monaghan [25]

$$\Pi_{ab} = \begin{cases} \dfrac{-\alpha(c_a + c_b)\mu_{ab} + 2\beta\mu_{ab}^2}{(\rho_a + \rho_b)}, & (\bm{u}_a - \bm{u}_b) \cdot (\bm{r}_a - \bm{r}_b) < 0 \\ 0, & (\bm{u}_a - \bm{u}_b) \cdot (\bm{r}_a - \bm{r}_b) \geq 0 \end{cases} \tag{25}$$

where

$$\mu_{ab} = \frac{h(\bm{u}_a - \bm{u}_b) \cdot (\bm{r}_a - \bm{r}_b)}{r_{ab}^2 + \eta}. \tag{26}$$

The parameters $\alpha$ and $\beta$ are used to control the strength of the artificial viscosity. $\alpha$ is related to the shear viscosity, and $\beta$ is related to the bulk viscosity.

For SPH simulation, the density and pressure fields may undergo large fluctuations numerically. In order to reduce the fluctuation, the Shephard filtering [27] is applied to reinitialize the density field.

$$\tilde{\rho}_a = \frac{\sum_b m_b W_{ab}}{\sum_b V_b W_{ab}} \tag{27}$$

In this paper, the summation is only executed for the particles from the same phase. The density reinitialization is conducted every 50 time steps for the liquid phase and every 500 time for the gas phase.

To prevent particle penetration, the XSPH correction introduced by Monaghan [25] is used to move

particles

$$\frac{d\mathbf{r}_a}{dt} = \hat{\mathbf{u}}_a = \mathbf{u}_a + \varepsilon \sum_b \frac{2m_b}{\rho_a + \rho_b}(\mathbf{u}_b - \mathbf{u}_a)W_{ab} \qquad (28)$$

Following Colagrossi and Landrini [28], the XSPH correction is also used in the mass equation.

**C. SPH formulations for interface**

For multiphase flow, especially for liquid-gas flow, there exists a discontinuity at the interface for certain fluid properties, such as density, viscosity and thermal conductivity. The discontinuity may lead to numerical difficulties. Therefore, the SPH equations for single phase fluid need to be modified for the liquid-gas interface.

Following Cleary and Monaghan [29], when two particles from different phases interact with each other, the following thermal conductivity is used.

$$\bar{\kappa}_{ab} = \frac{2\kappa_a \kappa_b}{\kappa_a + \kappa_b} \qquad (29)$$

Similarly, the viscosity between the gas and liquid particles is

$$\bar{\mu}_{ab} = \frac{2\mu_a \mu_b}{\mu_a + \mu_b}. \qquad (30)$$

For the pressure term, the inter-particle pressure proposed by Hu and Adams [30] is used to replace the particle pressure in Eq. (22) at the liquid-gas interface

$$\bar{p}_{ab} = \frac{\rho_a p_b + \rho_b p_a}{\rho_a + \rho_b}. \qquad (31)$$

The contributions of particle $b$ to the momentum equation and the energy equation of particle $a$ are as follows.

$$\frac{d\mathbf{u}_{ab}}{dt} = -m_b \left(2\frac{\bar{p}_{ab}}{\rho_a \rho_b} + \Pi_{ab} + R_{ab}\right)\nabla_a W_{ab}$$
$$+ \frac{2m_b \bar{\mu}_{ab}(\mathbf{r}_a - \mathbf{r}_b)\cdot\nabla_a W_{ab}}{\rho_a \rho_b (r_{ab}^2 + \eta)}(\mathbf{u}_a - \mathbf{u}_b) \qquad (32)$$

$$\frac{dT_{ab}}{dt} = \frac{2\bar{\kappa}_{ab} m_b (\mathbf{r}_a - \mathbf{r}_b)\cdot\nabla_a W_{ab}}{C_p \rho_a \rho_b (r_{ab}^2 + \eta)}(T_a - T_b) \qquad (33)$$

Here $R_{ab}$ on the right hand side of Eq. (32) is an artificial repulsive force with the following form

$$R_{ab} = -\varepsilon^R \left| \frac{\overline{p}_{ab}}{\rho_a \rho_b} \right| \tag{34}$$

The parameter $\varepsilon^R$ is in the range of 0 and 0.1. This repulsive force is similar to that used by Monaghan [31] and Grenier et al. [32].

When a gas particle interacts with a liquid particle, Eq. (21) tends to overestimate the contribution of the liquid particle to the density of the gas particle, because the mass of a liquid particle is much larger than the mass of a gas particle. In order to avoid the overestimation, the contribution of the liquid particle to the rate of change of the density of the gas particle is calculated by

$$\frac{d\rho_{gl}}{dt} = \rho_g V_l (\boldsymbol{u}_g - \boldsymbol{u}_l) \cdot \nabla_g W_{gl} \tag{35}$$

The subscripts g and l denote the gas particle and the liquid particle, respectively. The contribution of the gas particle to the rate of change of the density of the liquid particle is

$$\frac{d\rho_{lg}}{dt} = m_g (\boldsymbol{u}_l - \boldsymbol{u}_g) \cdot \nabla_l W_{lg}. \tag{36}$$

For the equation of the vapor mass fraction at the interface, a liquid particle is treated as a gas particle, and its vapor mass fraction is defined by the saturated vapor mass fraction, Eq. (14). The contribution of a liquid particle to the rate of change of the vapor mass fraction of a gas particle is

$$\frac{dY_{gl}}{dt} = \frac{2m_l D_g (\boldsymbol{r}_g - \boldsymbol{r}_l) \cdot \nabla_g W_{gl}}{\rho_l (r_{gl}^2 + \eta)} (Y_g - Y_l). \tag{37}$$

It should be noted that all the formulations in this section (i.e., Section III.C) are only used for the interface. That is, the interactions between two particles from different phases are calculated using the formulations in this section, while the interactions between particles from the same phase are calculated using the formulations in Section III.B.

### D. SPH formulations for evaporation rate

The rate of mass transfer from a liquid particle to a gas particle due to evaporation, Eq. (12), is discretized as

$$\dot{m}_{gl} = \frac{2m_g m_l D_g (\boldsymbol{r}_g - \boldsymbol{r}_l) \cdot \nabla_g W_{gl}}{\rho_l (r_{gl}^2 + \eta)(1 - Y_g)} (Y_g - Y_l). \tag{38}$$

The total mass change rate of a gas particle is

$$\frac{dm_g}{dt} = \sum_l \dot{m}_{gl} = \sum_l \frac{2 m_g m_l D_g (\mathbf{r}_g - \mathbf{r}_l) \cdot \nabla_g W_{gl}}{\rho_l (r_{gl}^2 + \eta)(1 - Y_g)} (Y_g - Y_l). \qquad (39)$$

The total mass change rate of a gas particle is

$$\frac{dm_l}{dt} = -\sum_g \dot{m}_{gl} = -\sum_g \frac{2 m_g m_l D_g (\mathbf{r}_g - \mathbf{r}_l) \cdot \nabla_g W_{gl}}{\rho_l (r_{gl}^2 + \eta)(1 - Y_g)} (Y_g - Y_l). \qquad (40)$$

Eqs. (39) and (40) indicate that the total mass of the liquid and gas particles does not change. Thus, the mass conservation is satisfied in the process of evaporation.

The volumetric mass flux, Eq. (13), is

$$\dot{m}_g''' = \frac{\dot{m}_g}{V_g} = \sum_l \frac{2 \rho_g m_l D_g (\mathbf{r}_g - \mathbf{r}_l) \cdot \nabla_g W_{gl}}{\rho_l (r_{gl}^2 + \eta)(1 - Y_g)} (Y_g - Y_l). \qquad (41)$$

**E. Particle splitting and merging**

The phase change due to evaporation will increase the mass of a gas particle and decrease the mass of a liquid particle at the interface. The mass change rate of a gas particle and a liquid particle are given by Eqs. (39) and (40), respectively. In order to ensure that the mass of a particle is not excessively large or small, particle splitting and merging techniques are developed here. Both the splitting and merging process satisfy the conservation of mass, momentum and energy.

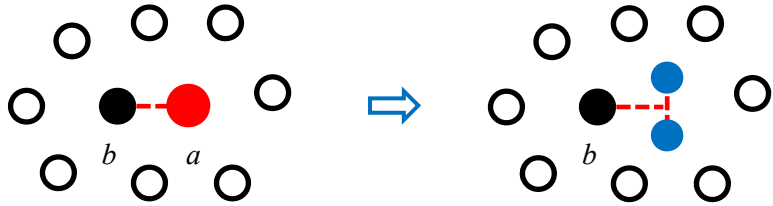

FIG. 2. Schematic of simulating particle splitting.

If the mass of a particle is larger than a given value, the particle will split into two smaller particles, as shown in Fig. 2. The process of particle splitting is as follows.

1) A reference mass is set to $m_r = \rho_r ds^d$, where $\rho_r$ is the reference density, $ds$ is the initial particle distance, and the superscript $d$ is the number of spatial dimension. In this study, a two-dimensional case is considered, thus $m_r = \rho_r ds^2$.

2) If the following condition is satisfied, particle $a$ will be split into two smaller particles.

$$m_a/m_r > \gamma_{max} \qquad (42)$$

Here $m_a$ is the mass of particle $a$. $\gamma_{max}$ is a parameter to control the maximum limit of particle mass, whose range is $1.5 \leq \gamma_{max} \leq 2$. Both the two smaller particles have the mass that is half of the mass of the original particle, and the same density and velocity of the original particle.

3) The next step is to find the nearest particle $b$ of particle $a$. The two new particles are on the perpendicular line of the line connecting particles $a$ and $b$. The distance between the two new particles is $\sqrt{m_a/\rho_r}/2$. The reason to find the nearest particle is to avoid that the new smaller particles are too close to the neighboring particles.

If the mass of a particle is less than a given value, it will merge to its nearest particle, as shown in Fig. 3. The process of particle splitting is as follows.

1) A reference mass is set to $m_r = \rho_r ds^2$.

2) If the following condition is satisfied, the particle $a$ will merge with its nearest particle.

$$m_a/m_r < \gamma_{min} \qquad (43)$$

Here $\gamma_{min}$ is a parameter to control the minimum limit of particle mass, whose range is $0.5 \leq \gamma_{min} < \gamma_{max}/2$. The reason to merge into the nearest particle is to avoid that the new particle is too close to the neighboring particles and to reduce to influence area of the merging process.

3) The next step is to find the nearest particle $b$ of particle $a$. The new particle is located at the center of mass of particles $a$ and $b$.

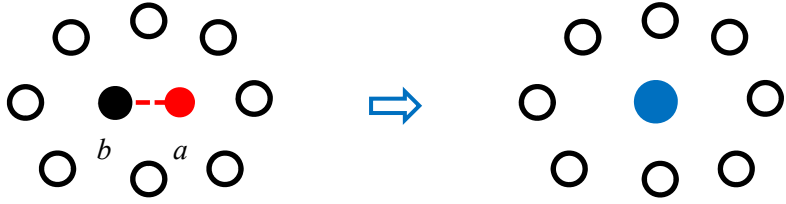

FIG. 3. Schematic of simulating particle merging.

## IV. NUMERICAL EXAMPLES

The evaporation model based on the SPH method will be validated in this section, by simulating three different cases. Table 1 shows the physical properties of the liquid and the gas used in the following numerical examples in this paper. Note that the density in the table is the initial density. The liquid density will change slightly during the simulation because the numerical method used in the paper is the so-called weakly compressible SPH method, which allows the density for up to a one percent variation from the initial density. On the other hand, the gas density does vary because of evaporation.

Table 1. Physical properties of the liquid and gas phases [33].

|  | $\rho$ (kg/m$^3$) | $\mu$ (kg/m/s) | $\kappa$ (W/m/K) | $C_p$ (J/kg/K) | $M$ (kg/mol) | $h_v$ (J/kg) | $D_v$ (m$^2$/s) | $T_B$ (K) |
|---|---|---|---|---|---|---|---|---|
| Gas | 1.2 | $2\times10^{-5}$ | 0.046 | 1000 | 0.029 |  | $2\times10^{-5}$ |  |
| Liquid | 1000 | $1\times10^{-3}$ | 0.6 | 4180 | 0.018 | $2.3\times10^6$ |  | 373 |

### A. The Stefan problem

To validate the new evaporation rate, Eq. (12), which was derived in this paper, and its SPH form, Eq. (38), the Stefan problem was simulated. As illustrated in Fig. 4, an open container was partially filled with liquid, and the remainder with gas. The liquid, then evaporates from the liquid-gas interface, and the vapor diffuses from the interface to the open end of the container. The vapor mass fraction at the interface is assumed to be constant (i.e., saturated vapor condition, $Y_{v,s}$), and the vapor mass fraction at the open end is also constant ($Y_{v,H}$). In other words, the system is at steady state, and the analytical solution of the vaporization mass flux is [21]

$$\dot{m}_v'' = \frac{\rho D_v}{H} \ln\left(\frac{1-Y_{v,H}}{1-Y_{v,s}}\right). \tag{44}$$

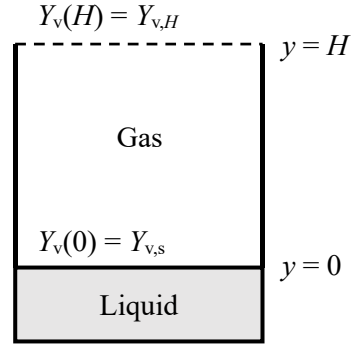

FIG. 4. Schematic of the Stefan problem.

Since the vapor mass fraction at the interface is assumed to be constant, and the interface is assumed to be stationary, the numerical simulation is only conducted in the gas phase. The bottom boundary of the computational domain is the liquid-gas interface, at which the vapor mass fraction is set from 0.1 to 0.9. The top boundary is a gas boundary, at which the vapor mass fraction is set at 0. The periodic boundary condition is used for the left and right boundaries. The height and width of computational domain are 2.0 mm and 0.5 mm, respectively. The initial SPH particle spacing is 0.05 mm. Fig. 5 shows the SPH results of evaporating mass flux compared with the analytical solution. The SPH prediction agrees well with the analytical solution.

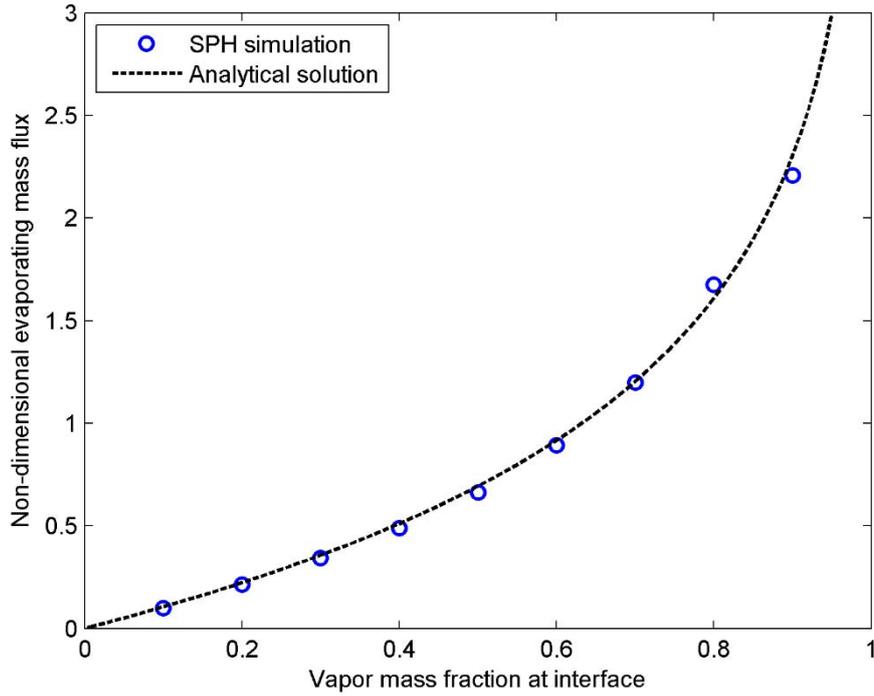

FIG. 5. SPH prediction and analytical solution of evaporating mass flux as a function of vapor mass fraction at the interface.

Another numerical test was conducted by solving only the equation for vapor mass fraction, Eq. (7), using SPH Eq. (24), without solving any other governing equations. The results are shown in Fig. 6. The numerical solution closely agreed with the analytical solution when the vapor mass fraction is less than 0.5. However, as the vapor mass fraction increased beyond 0.5, the numerical solution deviated from the analytical solution. According to Safari et al. [7], the divergence of the velocity at the liquid-gas interface is nonzero because of evaporation, which leads to the over-prediction of the evaporating mass flux. Therefore, the equation for vapor mass fraction, Eq. (7), alone does not accurately simulate the evaporation process. Therefore, for simulation of evaporation, all the governing equations listed in Section II need to be solved.

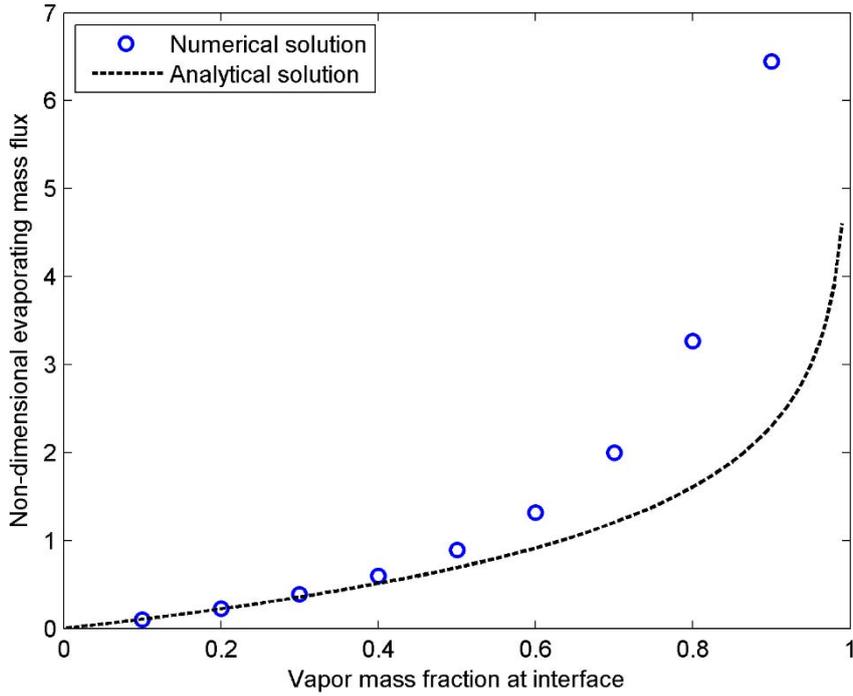

FIG. 6. Analytical solution and numerical prediction by considering only Eq. (24).

**B. Evaporation of a static drop**

The evaporation of a static drop was simulated using the proposed SPH method. Figure 7 shows the initial SPH particle distribution for simulating the evaporation of a static drop. The initial radius of the drop is $R_0 = 0.15$ mm. The initial temperature of the drop is 353 K. The drop was located at the center of a square computational domain, which was filled with gas. The length of the square was 1.2 mm. The initial temperature of the gas was 373 K. The temperature of the boundary was also 373 K, and did not change during the simulation. These temperatures were chosen in order to be consistent with and to allow comparisons with the conditions in the literature [6]. The initial vapor mass fraction in the gas phase was zero. The vapor mass fraction of the boundary remained zero. The initial particle spacing was 0.02 mm.

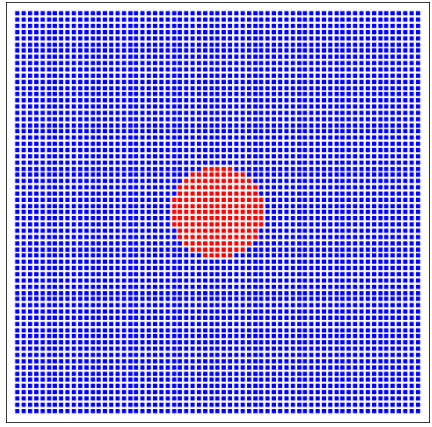

FIG. 7. Initial SPH particle distribution of a static drop, and the computational domain.

The interaction between the SPH particles along the interface was not absolutely symmetric. Thus, the shape of the interface was not a perfect circle, and the drop moved slightly. Although the movement of the drop was very slow, the drop had a noticeable displacement when time allowed. To avoid the movement, the drop was fixed at the center of the computational domain by use of the following equations.

$$\boldsymbol{r} = \boldsymbol{r} - \boldsymbol{r}_c, \quad \boldsymbol{u} = \boldsymbol{u} - \boldsymbol{u}_c \tag{45}$$

Here $\boldsymbol{r}_c$ and $\boldsymbol{u}_c$ are the displacement and velocity of the center of mass of the drop, respectively.

Figure 8 shows the snapshots of the evaporating drop at different times. The shape of the interface changed slightly with time, but it is very close to a circle. Figure 8 also shows that the size of the drop decreased slightly. The decrease in the drop size, as compared with the result from a 2D axisymmetric level set method [6], is shown in Fig. 9. It should be noted that the 2D circle used in this study corresponded to the cross section of a 3D cylinder of infinite length, while the 2D axisymmetric circle used in Ref. [6] corresponded to a 3D sphere. Therefore, the comparison in Fig. 9 qualitatively demonstrates the accuracy of the proposed SPH method. Since the ratio of surface area to volume of a 2D drop (this study) is less than that of a 2D axisymmetric drop (Ref. [6]), the decrease in the size of the 2D drop is less than that of the 2D axisymmetric drop, as shown in Fig. 9. Nonetheless, the trends are similar. At the initial stage, the size of the drop decreased quickly, because initially there was no vapor in the gas phase, and because the evaporation rate was fast. As the vapor concentration in the gas phase increased, the evaporation rate decreased.

As can be seen in Fig. 8, the SPH particle distribution was not uniform. The reason for this is that the sizes of the particles were not the same. As discussed in Section E, the ratio of the particle mass to the corresponding reference mass may vary from $\gamma_{min}$ to $\gamma_{max}$. Initially, the distribution of the particles was uniform, as shown in Fig. 7. Then the mass of the gas particles near the interface increased because of the mass transfer from the liquid particles to the gas particles due to evaporation. When the mass ratio of a gas particle was larger than $\gamma_{max}$, the particles were split into two smaller particles. At the same time, the mass of the liquid particles near the interface decreased. When the mass ratio of a liquid particle was less than $\gamma_{min}$, the liquid particle merged into its nearest liquid particle. The mass of the gas particles near the boundary also decreased because of the mass transfer from the gas particles to the boundary particles. When the mass ratio of a gas particle was less than $\gamma_{min}$, the gas particle merged into its nearest gas particle.

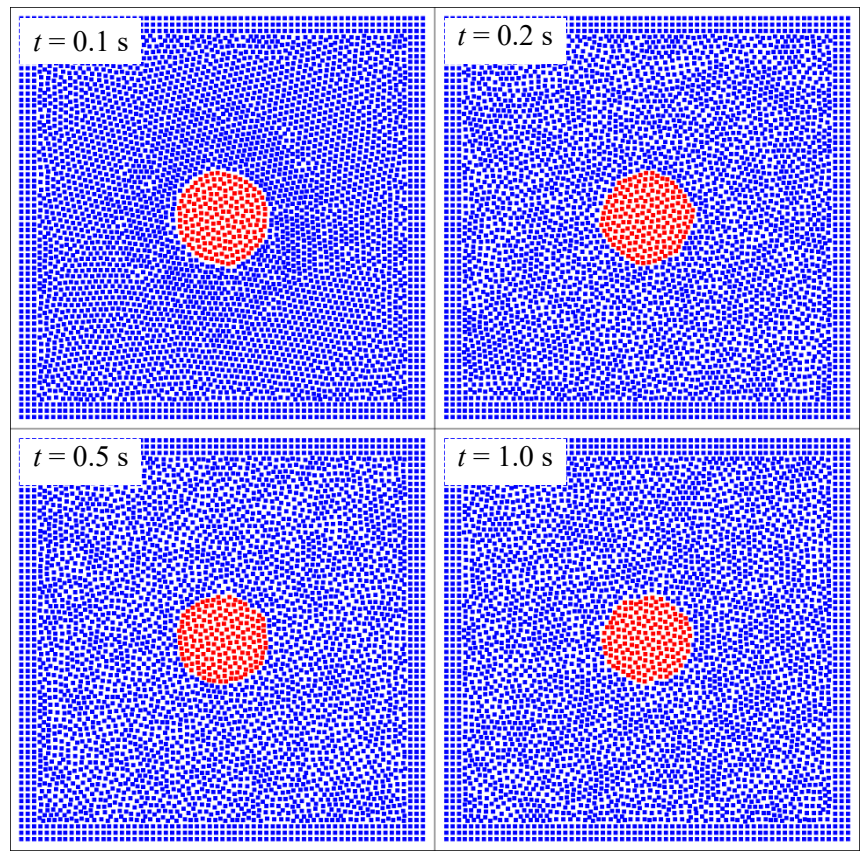

FIG. 8. Snapshots of the evaporating drop at different times.

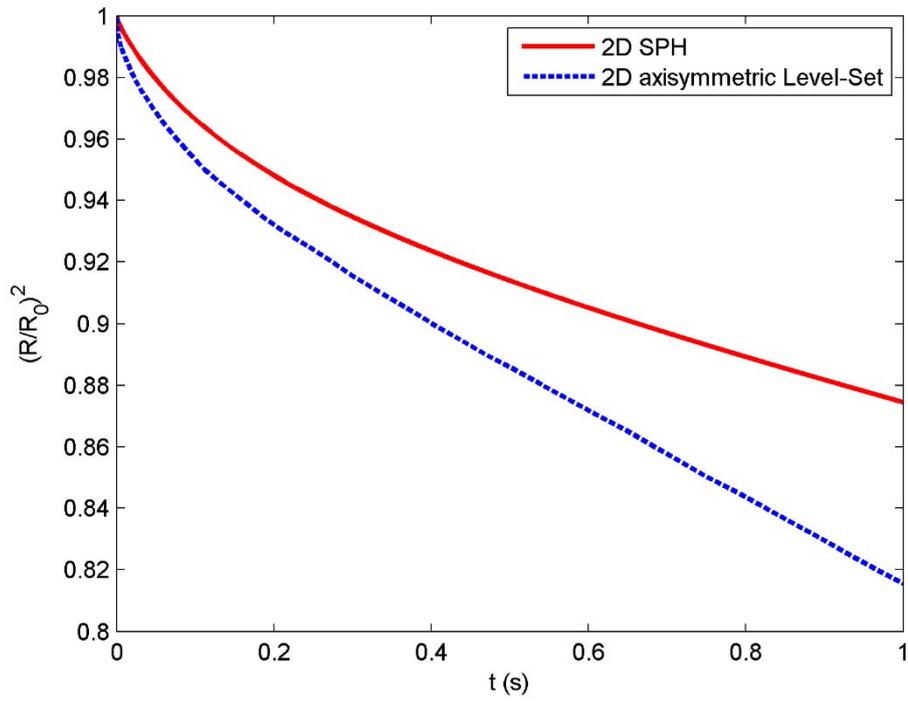

FIG. 9. Normalized square of radius versus time.

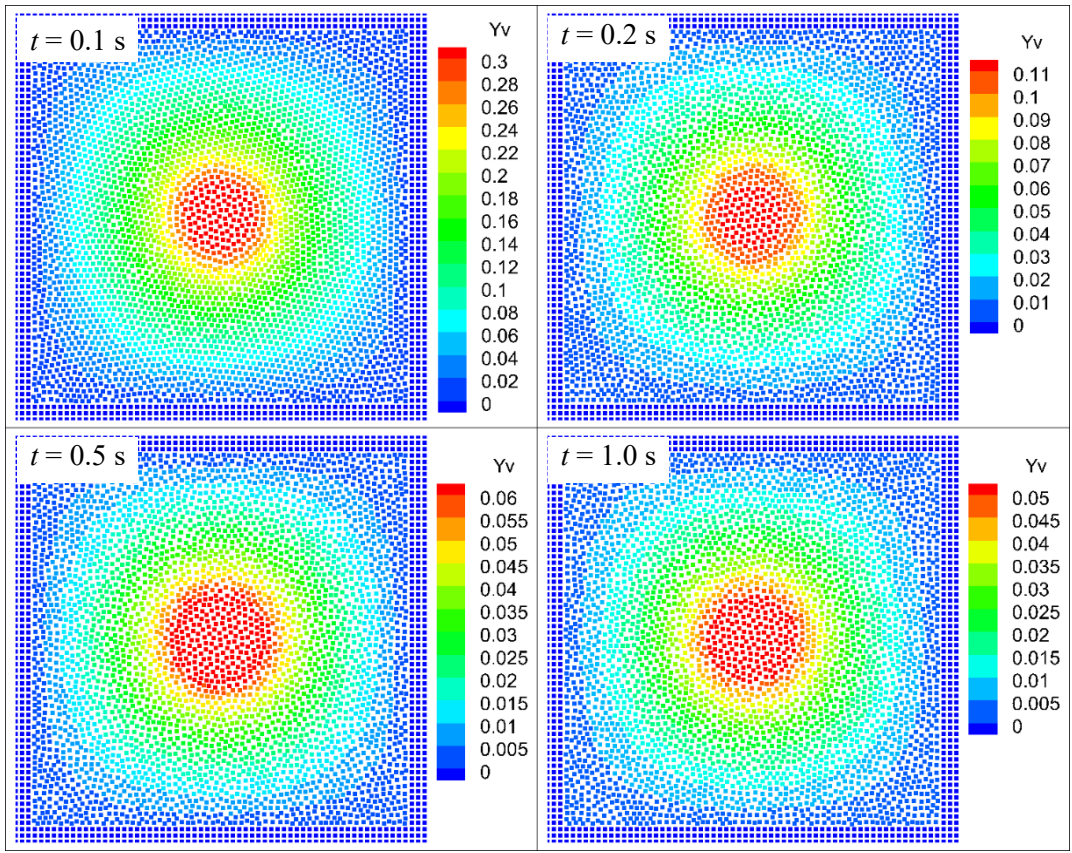

FIG. 10. Evolution of vapor mass fraction.

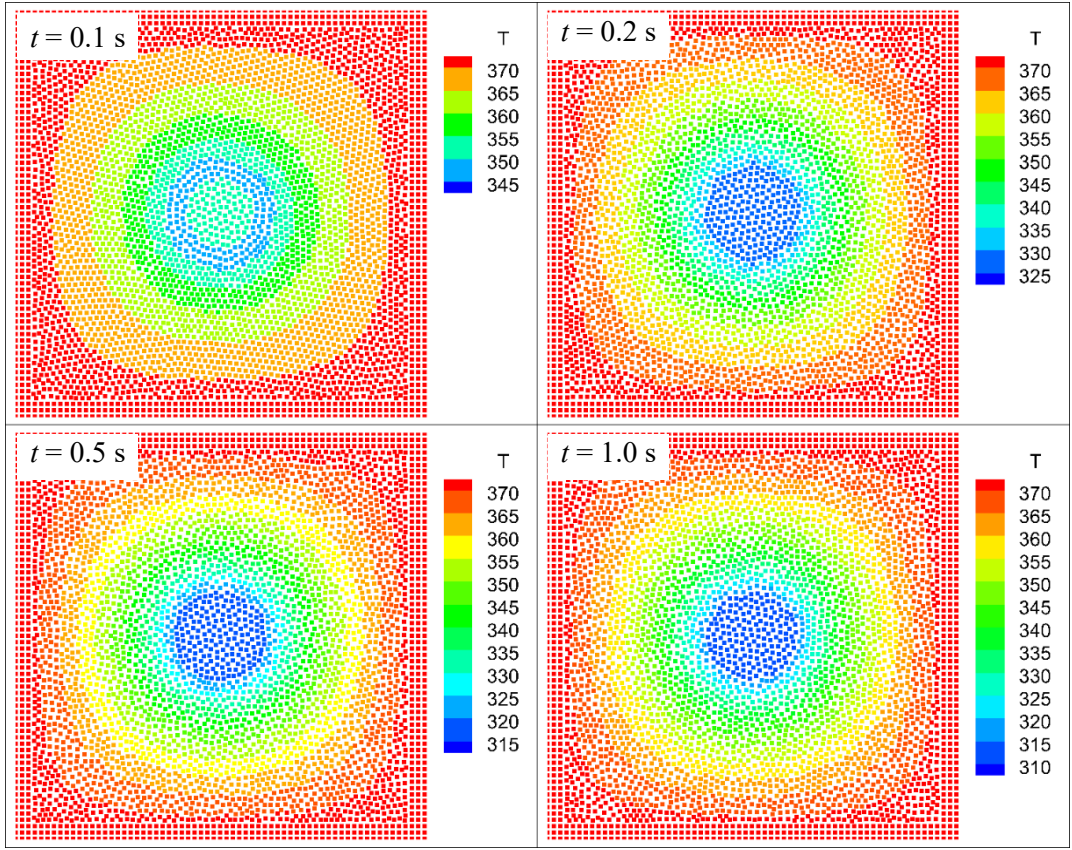

FIG. 11. Evolution of temperature.

Figure 10 shows the evolution of the vapor mass fraction surrounding the drop. As time increased from 0.1 s to 1.0 s, the corresponding saturated vapor mass fraction at the interface decreased from 0.3 to 0.05. The reason for this is that evaporation consumed energy, and thus decreased the drop temperature, and consequently decreased the vapor concentration at the interface. The evolution of the temperature field is shown in Fig. 11 to clearly show that the temperature of the drop decreased due to evaporation. Figure 11 also shows that the temperature of the drop was lower than its initial temperature, and that it decreased until reaching an equilibrium temperature. At certain times, the temperature at the interface was lower than the temperature at the drop center. Eventually, the temperature difference between the interface and the drop center decreased until reaching an equilibrium temperature. If the details of mass and energy transfer at the interface had not been considered, the temperature of the drop would have been higher than its initial temperature, and the temperature at the interface would have been higher than the temperature at the drop center, because the surrounding gas would have heated the liquid drop, as is commonly seen in traditional evaporation models.

## C. Evaporation of a drop impacting on a hot surface

The proposed method was also used to simulate the evaporation of a drop impacting on a hot surface, as shown in Fig. 12. The initial radius of the drop was $R = 0.25$ mm. The initial velocity of the drop was $U = 2$ m/s. The height and length of the computational domain were 1.5 mm and 5.0 mm, respectively. The drop was located at the center of the domain and surrounded by gas. The initial temperature of the drop was 353 K. The initial temperature of the gas was 373 K. The temperature of the boundaries was also 373 K, and did not change during the simulation. The initial vapor mass fraction in the gas phase was zero. The vapor mass fraction of the boundary remained zero. The initial particle spacing was 0.02 mm.

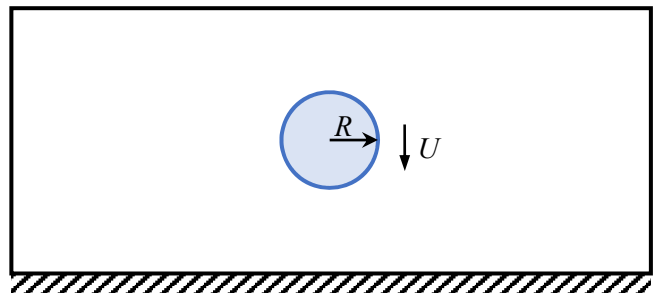

FIG. 12. Schematic of drop impact on a surface.

Figure 13 shows the evolution of drop impact on a hot surface. After the drop touched the surface, it spread and formed a film on the surface. At approximately 1.0 ms, a tiny crown-like structure was formed around the rim. Later, the crown merged with the film, and the film receded. Finally, the film reached an equilibrium size.

The evolution of the temperature field, and vapor mass fraction, are shown in Figs. 14 and 15, respectively. Since the initial temperature of the drop was lower than the gas temperature, the heat transfer from the surrounding gas to the drop led to the decrease in the local gas temperature. However, the drop temperature also decreased slightly because evaporation consumed energy, as discussed earlier. As can be seen in Fig. 14 ($t = 1.0$ and 2.0 ms), the rim had the lowest temperature, because the evaporation rate in the area is large. When the drop spreads on the hot surface and forms a film, heat transfer from the hot surface to the film increased the temperature of the film.

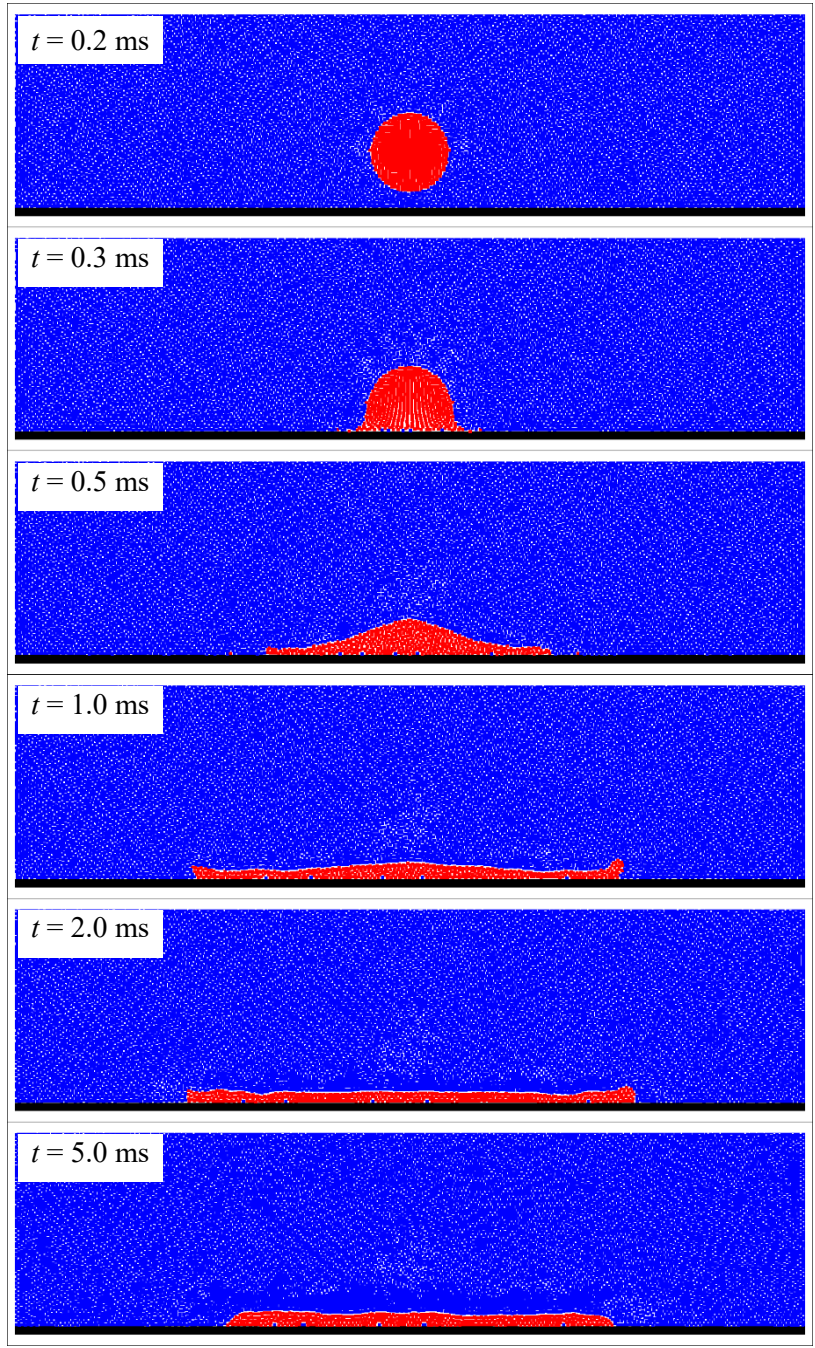

FIG. 13. Evolution of drop impact on a hot surface.

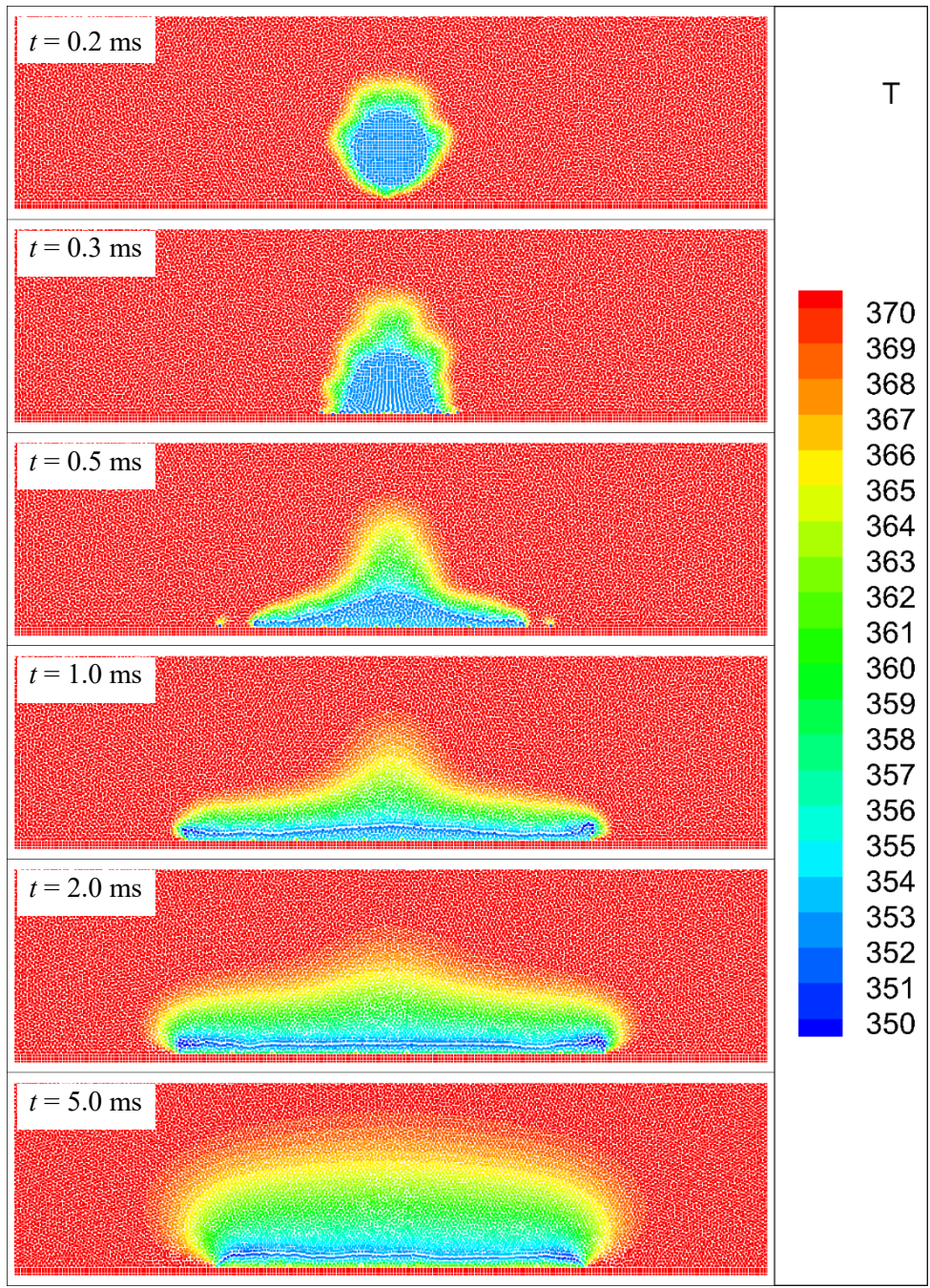

FIG. 14. Evolution of temperature field of drop impact on a hot surface.

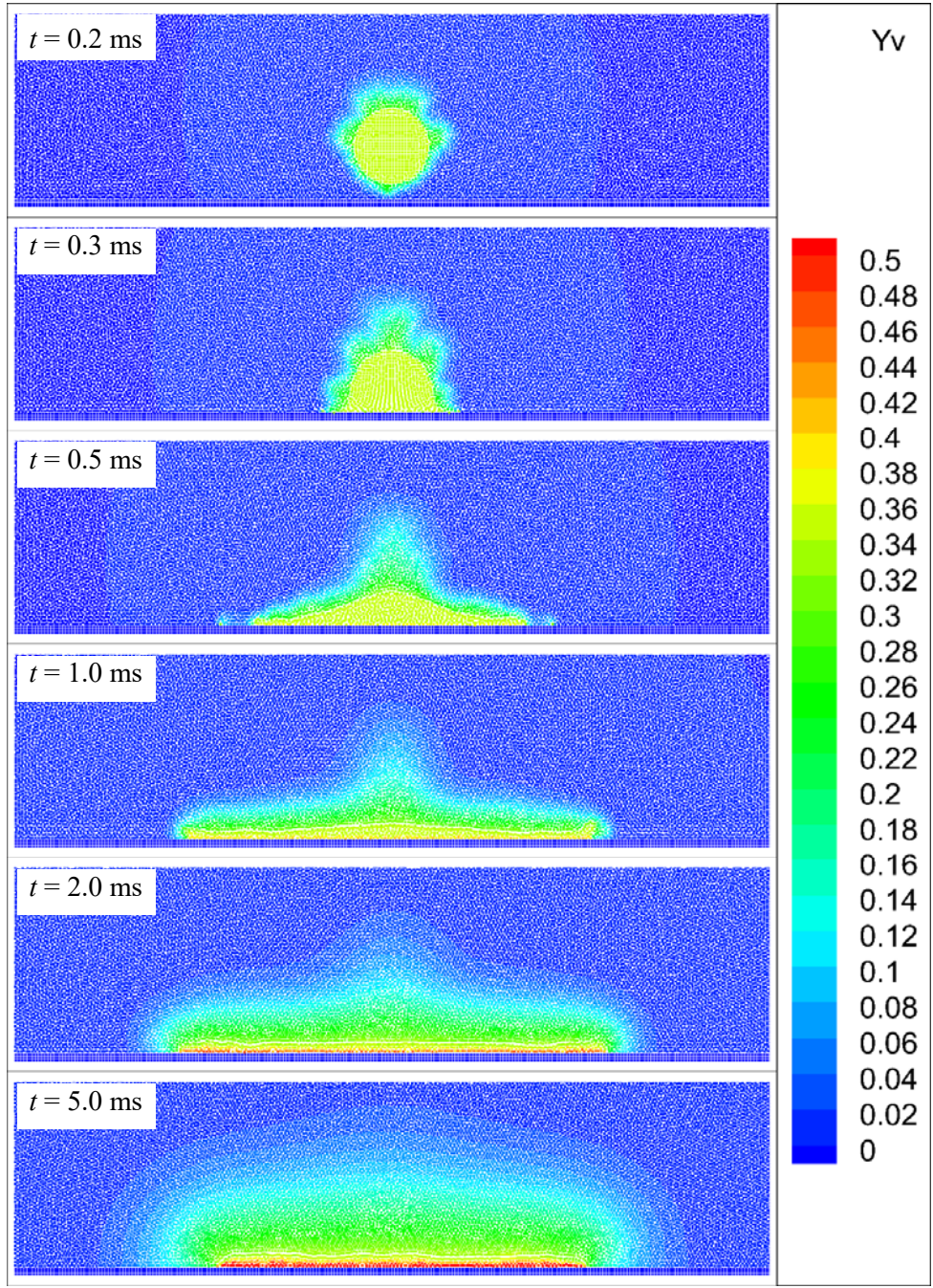

FIG. 15. Evolution of the vapor mass fraction of drop impact on a hot surface.

## V. CONCLUSION

The intent of this paper was to present an SPH method to simulate evaporating multiphase flows. This method accurately models the process of evaporation at the liquid-gas interface and the diffusion of the vapor species in the gas phase. An evaporating mass rate was derived to calculate the mass transfer at the interface. To model the process of phase change from the liquid phase to the gas phase, mass was allowed to transfer from a liquid SPH particle to a gas SPH particle. Thus this proposed method, unlike the traditional SPH method, allows change in the mass of an SPH particle. Additionally, particle splitting and merging techniques were developed to avoid the large difference in the SPH particle mass.

Three numerical examples were tested and compared with analytical solutions and results from a level-set method. In general, the results show that the method proposed in this paper successfully replicated the physical process of evaporating flows, such as heat and mass transfers and the diffusion of the vapor species. The first example were the Stefan problem, in which the mass evaporation rates at different conditions were predicted; the numerical results showed that the evaporation rate increased quickly as the vapor mass fraction at the interface increased, and that the results agree well with the analytical solution. The second example was to simulate the evaporation of a static drop—because of evaporation, the present SPH method predicts the decreases of both the temperature of the interface and the size of the drop. The last example was to simulate the evaporation of a drop impacting a hot surface. The temperature of the liquid-gas interface decreased at first because of evaporation, especially at the rim of the film. Then the temperature increased because of the heat transfer from the hot surface to the liquid. In summary, the results of this study indicate that the numerical method proposed in this paper can be successfully used to produce an evaporating flow simulation.

**References**

[1] W. A. Sirignano, *Fluid dynamics and transport of droplets and sprays* (Cambridge University Press, 2014), Third edn.
[2] K. Harstad and J. Bellan, Combustion and flame **137**, 163 (2004).
[3] L. Zhang and S.-C. Kong, Combustion and Flame **158**, 1705 (2011).
[4] L. Zhang and S.-C. Kong, Combustion and Flame **157**, 2165 (2010).
[5] S. S. Sazhin, Progress in energy and combustion science **32**, 162 (2006).
[6] S. Tanguy, T. Ménard, and A. Berlemont, J. Comput. Phys. **221**, 837 (2007).
[7] H. Safari, M. H. Rahimian, and M. Krafczyk, Phys. Rev. E **90**, 033305 (2014).
[8] H. Safari, M. H. Rahimian, and M. Krafczyk, Phys. Rev. E **88**, 013304 (2013).
[9] N. Nikolopoulos, A. Theodorakakos, and G. Bergeles, Int. J. Heat Mass Tran. **50**, 303 (2007).
[10] G. Strotos, M. Gavaises, A. Theodorakakos, and G. Bergeles, Int. J. Heat Mass Tran. **51**, 1516 (2008).
[11] L. B. Lucy, Astron. J. **82**, 1013 (1977).
[12] R. A. Gingold and J. J. Monaghan, Mon. Not. R. Astron. Soc. **181**, 375 (1977).
[13] J. J. Monaghan, Eur. J. Mech. B/Fluid **30**, 360 (2011).
[14] M. B. Liu and G. R. Liu, Arch. Comput. Methods Eng. **17**, 25 (2010).

[15] S. Li and W. K. Liu, Appl. Mech. Rev. **55**, 1 (2002).
[16] S. Nugent and H. A. Posch, Phys. Rev. E **62**, 4968 (2000).
[17] L. D. G. Sigalotti, J. Troconis, E. Sira, F. Peña-Polo, and J. Klapp, Phys. Rev. E **92**, 013021 (2015).
[18] M. Ray, X. Yang, S.-C. Kong, L. Bravo, and C.-B. M. Kweon, P. Combust. Inst. **36**, 2385 (2017).
[19] A. Das and P. Das, J. Comput. Phys. **303**, 125 (2015).
[20] P. W. Cleary, Appl. Math. Model **22**, 981 (1998).
[21] S. R. Turns, *An Introduction to Combustion: Concepts and Applications* (McGraw Hill, New York, 2012), Third edn.
[22] X. Yang, M. Liu, and S. Peng, Comput. Fluids **92**, 199 (2014).
[23] X.-F. Yang and M.-B. Liu, Acta Phys. Sin. **61**, 224701 (2012).
[24] J. W. Swegle, D. L. Hicks, and S. W. Attaway, J. Comput. Phys. **116**, 123 (1995).
[25] J. J. Monaghan, Ann. Rev. Astron. Astrophys. **30**, 543 (1992).
[26] J. J. Monaghan, Rep. Prog. Phys. **68**, 1703 (2005).
[27] J. Bonet and T.-S. Lok, Comput. Methods Applied Mech. Engrg. **180**, 97 (1999).
[28] A. Colagrossi and M. Landrini, J. Comput. Phys. **191**, 448 (2003).
[29] P. W. Cleary and J. J. Monaghan, J. Comput. Phys. **148**, 227 (1999).
[30] X. Y. Hu and N. A. Adams, J. Comput. Phys. **227**, 264 (2007).
[31] J. J. Monaghan, J. Comput. Phys. **159**, 290 (2000).
[32] N. Grenier, M. Antuono, A. Colagrossi, D. Le Touzé, and B. Alessandrini, J. Comput. Phys. **228**, 8380 (2009).
[33] S. M. Hosseini and J. J. Feng, Chem. Eng. Sci. **64**, 4488 (2009).